\documentclass[twocolumn, eqsecnum, aps]{revtex4}
\usepackage{graphicx}
\usepackage{dcolumn}

\begin{document}
%\input psbox 
%\baselineskip = .5\baselineskip  % single space

\title{A New Limit for the Coupling Constant of the Electron-Nucleus Scalar-Pseudoscalar Interaction} 
\vspace*{0.5cm}

\author{$^{1,2}$Bijaya K. Sahoo \protect \footnote[2] {E-mail: bijaya@iiap.res.in, B.K.Sahoo@gsi.de}, $^{1}$Rajat K. Chaudhuri, $^{1}$B. P. Das,
 $^{3}$Debashis Mukherjee, $^{4}$E. P. Venugopal \\
\vspace{0.3cm}
$^{1}${\it Non-Accelerator Particle Physics Group,\\Indian Institute of 
Astrophysics, Bangalore-560 034, India} \\
$^{2}${\it Atomphysik,\\Gesellschaft f\"ur Schwerionenforschung mbH, Planckstr.1, 64291 Darmstadt, Germany}\\
$^{3}${\it Department of Physical Chemistry, \\Indian Association for Cultivation of Science, Calcutta-700 032, India}\\
$^{4}${\it Department of Chemistry and Biochemistry \\ University of Detroit, Mercy, Detroit MI 48221, USA}}
%\email{bijaya@iiap.res.in,B.K.Sahoo@gsi.de}
\date{\today}

\begin{abstract}
\noindent
 We report the results of our calculations of the atomic
 electric dipole moments of cesium and thallium arising from 
 the electron-nucleus scalar-pseudoscalar interaction. The
 calculations are based on the all order relativistic coupled-cluster
 theory. Electron correlation effects, particularly in
 the case of thallium are of crucial importance. We obtain a
 new limit for the scalar-pseudoscalar interaction by combining
 the result of our thallium calculation and the measured value
 of the electric dipole moment of that atom.
\end{abstract} 
\maketitle

%\noindent
It is now widely recognized that atomic electric dipole moments (EDMs) arising from the violations of parity (P) and time-reversal (T) symmetries can provide important information about new physics beyond the Standard Model (SM) \cite{barr}. T violation implies CP violation via the CPT theorem \cite{luders}. The dominant sources of the EDM of a paramagnetic atom are the EDM of an electron and the scalar-pseudoscalar (S-PS) interaction between the electron and the nucleus which violates P as well as T symmetries \cite{shukla}. While the former has been dealt with extensively \cite{bernreuther}, the latter has received relatively little attention. Barr has shown that in two Higgs doublet models \cite{barr1}, the contribution of the S-PS interaction to an atomic EDM can exceed that due to the EDM of an electron for certain values of the model parameters. The S-PS electron-nucleus interaction
which originates from the S-PS electron-quark interactions could
therefore shed light on new physics beyond the Standard Model. 

In this Letter we calculate the ratios of the atomic EDMs to the S-PS coupling constants for cesium (Cs) and thallium (Tl) using relativistic coupled-cluster (RCC) theory. Combining the result of our Tl calculation with experimental data we obtain a new limit for the S-PS coupling constant ($C_S$). Following the EDM measurement of atomic Cs by Murthy {\it et al.} \cite{murthy}, a series of experiments were carried out by Commins and co-workers on atomic Tl \cite{abdullah,commins,regan}. Currently the most accurate data on the measurement of the EDM of any paramagnetic atom comes from the experiment on Tl by Regan {\it et al.} \cite{regan}. New experiments on the EDM of atomic Cs using the techniques of laser cooling and trapping are in progress in three different laboratories \cite{chin,weiss,heinzen,heinzen1}. They could improve the existing limit of the atomic EDM by about two orders of magnitude.

The P and T violating S-PS electron-nucleon interaction Hamiltonian ($\text{H}_{\text{EDM}}^{\text{S-PS}}$) is given by
\begin{eqnarray}
\text{H}_{\text{EDM}}^{\text{S-PS}} &=& \frac {G_F}{\sqrt{2}} C_S A \sum_e i\beta_e \gamma_e^5 \rho_N(r_e).
\end{eqnarray}
In the above expression $G_F$ is the Fermi constant, $C_S$ is the dimensionless S-PS constant which is given by $C_S = (Z C_{S,p} + N C_{S,n})/A$, where $Z$, $N$ and  $A$ are the atomic, neutron and atomic mass numbers respectively. It is a weighted average of the electron-neutron and electron-proton coupling constants, $C_{S,n}$ and $C_{S,p}$ with $C_{S,n} \approx C_{S,p}$. $\rho_N(r_e)$ is the nuclear density function and $\gamma^5(=i\gamma^0\gamma^1\gamma^2\gamma^3)$, which is a pseudo-scalar, is the product of the four Dirac matrices.

$\text{H}_{\text{EDM}}^{\text{S-PS}}$ is responsible for mixing states of opposite parities but with the same angular momentum. 
The strength of this interaction Hamiltonian is sufficiently
 weak for it to be considered as a first-order perturbation. It is therefore 
possible to write the atomic wave function as
\begin{equation}
|\Psi \rangle = |\Psi^{(0)} \rangle + G_F |\Psi^{(1)} \rangle .
\end{equation}
In RCC, the ground state $|\Psi_v^{(0)} \rangle$ for a single valence ($v$) open-shell system is given by \cite{lindgren,mukherjee}
\begin{eqnarray}
|\Psi_v^{(0)} \rangle = e^{T^{(0)}} \{1+S_v^{(0)}\} |\Phi_v \rangle
\end{eqnarray}
where we define $|\Phi_v \rangle= a_v^{\dagger}|\Phi_0\rangle$, with $|\Phi_0\rangle$ as the Dirac-Fock (DF) state for closed-shell system.

In the singles and doubles approximation we have
\begin{eqnarray}
T^{(0)} = T_1^{(0)} + T_2^{(0)} , \nonumber \\
S_v^{(0)} =  S_{1v}^{(0)} + S_{2v}^{(0)}
\end{eqnarray}
where $T_1^{(0)}$ and $T_2^{(0)}$ are the single and double particle-hole excitation operators for core electrons (holes) and $S_{1v}^{(0)}$ and $S_{2v}^{(0)}$ are the single and double excitation operators with the valence electron respectively.
The amplitudes corresponding to these operators can be determined by solving the RCC singles and doubles equations \cite{merlitz}. A subset of important triple excitations have been considered in the determination of the open shell amplitudes $S_{1v}^{(0)}$ and $S_{2v}^{(0)}$ which is described in \cite{kaldor}.

Using eqn. (0.2) the EDM of an atom of the ground state is given by
\begin{eqnarray}
D_A &=& \frac {\langle \Psi_v| \text{D} |\Psi_v \rangle } {\langle \Psi_v|\Psi_v \rangle } \nonumber \\
&=& \frac {\langle \Psi_v^{(0)}| \text{D} |\Psi_v^{(1)} \rangle + \langle \Psi_v^{(1)}| \text{D} |\Psi_v^{(0)} \rangle } {\langle \Psi_v^{(0)}|\Psi_v^{(0)} \rangle }
\end{eqnarray}
where higher order perturbed terms are neglected and D is the electric dipole (E1) operator. Using the explicit expression for the first order perturbed wave function ($|\Psi_v^{(1)} \rangle$), we get
\begin{eqnarray}
D_A &=& \sum_{I \ne v} \frac {\langle \Psi_v^{(0)}| \text{D} |\Psi_I^{(0)} \rangle \langle \Psi_I^{(0)}| \text{H}_{\text{EDM}}^{\text{S-PS}} |\Psi_v^{( 0)} \rangle } {\langle \Psi_v^{(0)}|\Psi_v^{(0)}\rangle(E_v - E_I)} \nonumber \\
&& + \sum_{I \ne v} \frac {\langle \Psi_v^{(0)}| \text{H}_{\text{EDM}}^{\text{S-PS}} |\Psi_I^{(0)} \rangle \langle \Psi_I^{(0)}| \text{D} |\Psi_v^{(0)} \rangle } {(E_v - E_I)\langle \Psi_v^{(0)}|\Psi_v^{(0)}\rangle} \\ 
 &=& \frac{2}{\langle \Psi_v^{(0)}|\Psi_v^{(0)}\rangle} \sum_{I \ne v} \frac {\langle \Psi_v^{(0)}| \text{D} |\Psi_I^{(0)} \rangle \langle \Psi_I^{(0)}| \text{H}_{\text{EDM}}^{\text{S-PS}} |\Psi_v^{( 0)} \rangle } {E_v - E_I} \nonumber  
\end{eqnarray}
since both D and $\text{H}_{\text{EDM}}^{\text{S-PS}}$ are hermitian. In the above equation $I$ represent intermediate states.

It is obvious from the above equation that, the accuracy of the calculation of 
$D_A$ depends on the excitation energies of the different intermediate states, the matrix elements of $\text{H}_{\text{EDM}}^{\text{S-PS}}$ and D.

In the present work we obtain the first order wave function as the solution of the equation 
\begin{eqnarray}
(\text{H}^{(0)} - E^{(0)})|\Psi_v^{(1)}\rangle = (E^{(1)} - \text{H}_{\text{EDM}}^{\text{S-PS}}) |\Psi_v^{(0)}\rangle 
\end{eqnarray}
where $E^{(1)}$ vanishes for first order correction. In this approach,  $|\Psi_v^{(1)}\rangle$ implicitly contains all the intermediate states.

The perturbed cluster operators can be written as
\begin{eqnarray}
T = T^{(0)} + G_F T^{(1)} , \nonumber \\
S_v =  S_v^{(0)} + G_F S_v^{(1)}
\end{eqnarray}
where $T^{(1)}$ and $S_v^{(1)}$ are the first order of $G_F$ corrections to the cluster operators  $T^{(0)}$ and $S_v^{(0)}$ respectively. The amplitudes of these operators are solved, keeping up to linear in EDM perturbed amplitudes, by the following equations
\begin{eqnarray}
\noindent
\langle \Phi_a^p |\overline{\text{H}_N^{(0)}} T^{(1)} + \overline{\text{H}_{\text{EDM}}^{\text{S-PS}}}| \Phi_0 \rangle &=& 0 , \nonumber\\
\langle \Phi_{ab}^{pq} |\overline{\text{H}_N^{(0)}} T^{(1)} + \overline{\text{H}_{\text{EDM}}^{\text{S-PS}}}| \Phi_0 \rangle &=& 0 ,
\end{eqnarray}
\noindent
and
\begin{eqnarray}
\langle\Phi_v^p|\overline{\text{H}_N^{(0)}} S_v^{(1)}+(\overline{H_N^{(0)}}T^{(1
)}+\overline{\text{H}_{\text{EDM}}^{\text{S-PS}}})\{1+S_v^{(0)}\}|\Phi_v\rangle \nonumber \\ =  - \langle\Phi_v^p|S_v^{(1)}|\Phi_v\rangle \text{IP} , \nonumber\\
\langle\Phi_{vb}^{pq}|\overline{\text{H}_N^{(0)}} S_v^{(1)}+(\overline{\text{H}_N^{(0)}
}T^{(1)}+\overline{\text{H}_{\text{EDM}}^{\text{S-PS}}})\{1+S_v^{(0)}\}|\Phi_v\rangle \nonumber \\ =  - \langle\Phi_{vb}^{pq}|S_v^{(1)}|\Phi_v\rangle \text{IP} , \
\end{eqnarray}
where $\text{H}^{(0)}$ is the Dirac-Coulomb (DC) Hamiltonian and $\overline{\text{H}}$ is defined as $e^{-T^{(0)}}\text{H}e^{T^{(0)}}$ which is computed after 
determining $T^{(0)}$, IP is the ionization potential corresponding to the
 valence electron '$v$' and the subscript $N$ denotes the normal form of an operator.
We have used $a,b..$ and $p,q..$ etc. to represent holes and particles respectively. $|\Phi_v^p\rangle$ and $|\Phi_{vb}^{pq}\rangle$ are the single and double
 excited states respectively with respect to the $|\Phi_v\rangle$. Using eqns.
(0.3), (0.5), (0.8) and only keeping terms linear in $G_F$, the expression for 
$D_A$ can be written as
\begin{table*}[top]
\caption{Excitation energies ($cm^{-1}$), E1 elements (a.u.) and magnetic dipole hyperfine structure constants (MHz) of important intermediate states in cesium and thallium. } 
\begin{ruledtabular}
\begin{tabular}{l|ccc|ccc}
 & \text{Cesium}  &  &  & \text{Thallium} & &  \\
\hline
\hline
Initial state & $6s_{1/2} \rightarrow 6p_{1/2} $ & & $6s_{1/2} \rightarrow 7p_{1/2} $ & $6p_{1/2} \rightarrow 7s_{1/2}$ & & $6p_{1/2} \rightarrow 8s_{1/2}$ \\
$\rightarrow$ Final state &  & & & & & \\
\hline
Excitation &  &  & & & & \\
energy & 11229.38 & & 21796.31 & 26038.62 & & 10462.32 \\
Experiment & 11177.84$^{a,b}$ & & 21765.30$^{a,b}$ & 26477.50$^a$ & & 10520.01$^a$ \\
\hline
E1 transition &  & &  & & &\\
amplitude & 4.53 & & 0.292 & 1.84 & & 0.57 \\
Experiment & 4.52(1)$^c$ & & 0.284(2)$^c$ & 1.81(2)$^e$ &  & - \\
\hline
\hline
Atomic state & $6s_{1/2}$ & $6p_{1/2}$ & $7p_{1/2}$ & $6p_{1/2}$ & $7s_{1/2}$ & $8s_{1/2}$ \\
 &  & & & & &\\
\hline
Hyperfine &  & & & & &\\
constant(A) & 2292.32 & 284.86  & 94.67  & 21025.98 & 11992.11 & 4118.57 \\
Experiment & 2298.16$^d$ & 291.90(9)$^d$ & 94.35(4)$^d$ & 21311$^e$ & 12297$^e$ & - \\
\end{tabular}
\end{ruledtabular}
References: $^a$ \cite{moore}; $^b$ \cite{weber}; $^c$ \cite{shabanova}; $^d$ \cite{arimondo}; $^e$ \cite{hsieh,gallagher}
\end{table*}
\begin{widetext}
\begin{eqnarray}
D_A = \frac {<\Phi_v |\{ 1+ S_v^{(1)^{\dagger}} + T^{(1)^{\dagger}}
S_v^{(0)^{\dagger}} + T^{(1)^{\dagger}} \} e^{T^{(0)^{\dagger}}} \text{D} e^{T^{
(0)}} \{ 1+ T^{(1)} + T^{(1)} S_v^{(0)} + S_v^{(1)} \} |\Phi_v > }
{1+N_v^{(0)}} \ \ \ \ \ \ \ \ \ \ \ \ \ \ \ \ \ \ \ \ \nonumber \\
=\frac {<\Phi_v|S_v^{(1)^{\dagger}}\overline{\text{D}^{(0)}}(1+S_v^{(0)})+(1+S_v^{(0)^{\dagger}})\overline{\text{D}^{(0)}}S_v^{(1)}+S_v^{(0)^{\dagger}} (T^{(1)^{\dagger}}\overline{\text{D}^{(0)}}+\overline{\text{D}^{(0)}}T^{(1)}) S_v^{(0)}+(T^{(1)^{\dagger}}\overline{\text{D}^{(0)}}+\overline{\text{D}^{(0)}}T^{(1)})S_v^{(0)}|\Phi_v>}{1+N_v^{(0)}} . \ \ \ \
\end{eqnarray}
\end{widetext}
In the above expression we define $\overline{\text{D}^{(0)}} = e^{T^{{(0)}^\dagger}}\text{D}e^{T^{(0)}}$ and $N_v^{(0)} = S_v^{{(0)}^{\dagger}}e^{T^{{(0)}^\dagger}}e^{T^{(0)}}S_v^{(0)}$ for the valence electron '$v$' and each term is connected.

The orbitals are constructed as linear combinations of Gaussian type orbitals (GTOs) of the form \cite{rajat}
\begin{equation}
F_{i,k}(r) = r^k e^{-\alpha_ir^2}
\end{equation}
where $k=0,1,..$ for s,p,.. type orbital symmetries respectively. For the
exponents, we have used
\begin{equation}
\alpha_i = \alpha_0 \beta^{i-1} .
\end{equation}

\begin{figure}
\label{fig:goldstone}
\includegraphics[width=8.0cm]{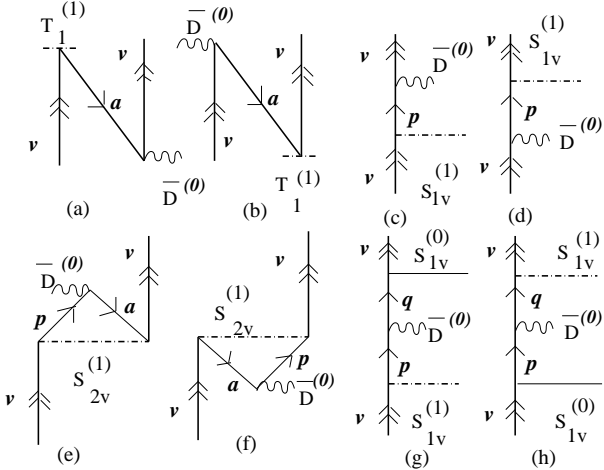}
\caption{Important Goldstone diagrams corresponding EDM amplitude calculations.}
\end{figure}
We have considered $\alpha_0$ as 0.00525 and
$\beta$ as 2.73 for all the symmetries for both the systems. To obtain Dirac-Fock (DF) wave function, we have undertaken 36$s_{1/2}$, 34$p_{1/2}$, 34$p_{3/2}$, 32$d_{3/2}$, 32$d_{5/2}$, 30$f_{5/2}$, 30$f_{7/2}$, 20$g_{7/2}$ and 20$g_{9/2}$ GTOs for cesium and 38$s_{1/2}$, 36$p_{1/2}$, 36$p_{3/2}$, 36$d_{3/2}$, 36$d_{5/2}$, 30$f_{5/2}$, 30$f_{7/2}$, 25$g_{7/2}$ and 25$g_{9/2}$ GTOs for thallium respectively. All core (occupied) electrons are excited for both the systems. All orbitals are generated on a grid \cite{rajat} using Fermi nuclear distribution \cite{parpia}.

We have presented in table I properties related to the EDM in the ground states of atomic Cs and Tl. The square root of the products of the relevant hyperfine constants (A) provide an estimate of accuracy of the EDM matrix elements. The results of the various properties of Cs and Tl are in most cases in very good agreement with the measured values.
\begin{table}[h]
\caption{Contributions from important RCC terms for $D_A/C_S$ calculations of $6s \ ^2S_{1/2}$ and $6p \ ^2P_{1/2}$ states for cesium and thallium respectively in $\times 10^{-18}e-cm$.} 
\begin{ruledtabular}
\begin{tabular}{l|ccc}
Important & \text{Cesium} & &\text{Thallium} \\
terms & \text{$6s \ ^2S_{1/2}^{(1)}$} & &\text{$6p \ ^2P_{1/2}^{(1)}$} \\
%  & $\times 10^{-18}e-cm$& & $ \times 10^{-18}e-cm$ \\
\hline
Contributions from DF & &  & \\
$\text{D} \text{H}_{\text{EDM}}^{\text{S-PS}} + \text{H}_{\text{EDM}}^{\text{S-PS}} \text{D}$ & \hfill-0.578 & & \hfill3.217 \\
\hline
 &  &  &\\
Contributions from RCC &  &  &\\
$\text{D} T_1^{(1)} + T^{(1)^{\dagger}} \text{D} $ & \hfill-0.035 & & \hfill3.056 \\
$\overline{\text{D}^{(0)}} S_{1v}^{(1)} + S_{1v}^{(1)\dagger} \overline{\text{D}^{(0)}}$ & \hfill-0.878 & & \hfill4.453 \\
$\overline{\text{D}^{(0)}} S_{2v}^{(1)} + S_{2v}^{(1)\dagger} \overline{\text{D}^{(0)}}$ & \hfill0.043 & & \hfill-3.835 \\
$S_{1v}^{(0)\dagger} \overline{\text{D}^{(0)}} S_{1v}^{(1)} + S_{1v}^{(1)\dagger} \overline{\text{D}^{(0)}} S_{1v}^{(0)} $ & \hfill0.015 & & \hfill-0.304 \\
$S_{2v}^{(0)\dagger} \overline{\text{D}^{(0)}} S_{1v}^{(1)} + S_{1v}^{(1)\dagger} \overline{\text{D}^{(0)}} S_{2v}^{(0)} $ & \hfill0.041  & & \hfill0.174 \\
$S_{1v}^{(0)\dagger} \overline{\text{D}^{(0)}} S_{2v}^{(1)} + S_{2v}^{(1)\dagger} \overline{\text{D}^{(0)}} S_{1v}^{(0)}$ & \hfill0.004 & & \hfill0.023 \\
$S_{2v}^{(0)\dagger} \overline{\text{D}^{(0)}} S_{2v}^{(1)} + S_{2v}^{(1)\dagger} \overline{\text{D}^{(0)}} S_{2v}^{(0)}$ & \hfill-0.008 & & \hfill-0.036 \\
Norm. & \hfill0.019 & & \hfill-0.032 \\
%\hline
 &  &  &\\
Total & \hfill-0.801 & & \hfill4.056 \\
\end{tabular}
\end{ruledtabular}
\end{table}

We have calculated the ratio of $D_A/C_S$ for Cs and Tl. Important Goldstone diagrams considered in these calculations are shown in FIG. 1 and the contributions from significant RCC terms are given in table II. The leading contributions for both Cs and Tl come from $DS_{1v}^{(1)}$ (diagrams 1((c)+(d))). These diagrams represent the DF, important pair correlation and a sub class of core-polarization effects. The DF contributions are -0.578 and 3.217 in the units given in the table for Cs and Tl respectively. Two different types of core-polarization effects represented by the terms $\text{D}T_1^{(1)}$ (diagrams 1((a)+(b))) and $DS_{2v}^{(1)}$ (diagrams 1((e)+(f)+(g)+(h))) make very large contributions but with opposite signs in the case of Tl. This is primarily due to the strong Coulomb and S-PS interactions between the $6p_{1/2}$ valence and the $6s_{1/2}$ core electron. However, these interactions are much weaker in the case of the $6s_{1/2}$ valence and the core electrons for Cs. The corresponding core-polarization contribution is therefore much smaller in size for Cs. It is of interest to note that the contributions of two other correlation effects represented by the terms $S_{1v}^{(0)\dagger}DS_{1v}^{(1)}$ and $S_{2v}^{(0)\dagger}DS_{1v}^{(1)}$ are non-negligible.

The results
of various calculations of $D_A/C_S$ are presented in 
table III. We have estimated the error by taking the difference 
between our RCC calculations with single, double and the leading
triple excitations and just single and double excitations. Bouchiat's calculation is based on an one electron potential and relativistic corrections are added to the wave functions \cite{bouchiat}. Venugopal's result for Cs is obtained using a hybrid method which combines certain important features of many-body perturbation theory and the multi-configurational Dirac-Fock method \cite{venugopal}. M{\aa}rtensson-Pendrill and Lindroth have computed the Dirac-Fock and various types of core-polarization effects \cite{martensson}. However, they have not considered pair-correlation effects which are included to all orders of the Coulomb interaction and one order in S-PS interaction in our work.

We have obtained a new limit for the S-PS coupling constant
$C_S=(0.995 \pm 1.756 \pm 0.033) \times 10^{-7}$ by combining our RCC calculation 
of $D_A/C_S$ for Tl and the measured value of $D_A$ for that atom \cite{regan};
the first uncertainty comes from the experiment and the second
from theory. This is a significant improvement over the current
limit $C_S= (2 \pm 7 ) \times 10^{-7}$ \cite{martensson}. It would be possible to improve this
limit even further if the new generation of EDM experiments using
cold cesium atoms reach their expected levels of accuracies \cite{chin,weiss,heinzen}. 
The results of these experiments could then be combined
with our calculation of $D_A/C_S$ for cesium (0.5\% accuracy) to
yield this new limit.
\begin{table}[t]
\caption{Comparison of $D_A/C_S$ results with others in $\times 10^{-18}e-cm$.}
\begin{ruledtabular}
\begin{tabular}{lccc}
 & Cesium &  & Thallium  \\
 & $D_A/C_S$  & & $D_A/C_S$  \\
\hline
Bouchiat \cite{bouchiat} & -0.689 & & - \\
Venugopal \cite{venugopal} & -0.805 & & - \\
M{\aa}rtensson-Pendrill &  &  & \\
and Lindroth \cite{martensson} & -0.72(1$\pm$0.03) &  & 7$\pm$2 \\
Present & -0.801($\pm$0.004) &  & 4.056($\pm$0.137) \\
\end{tabular}
\end{ruledtabular}
\end{table}

In summary, we have calculated $D_A/C_S$ for 
Cs and Tl using RCC theory with accuracies of 0.5\% and 3.3\%
respectively. Many-body effects were found to be of crucial
importance, particularly for Tl. Our calculated value of $D_A/C_S$
for Tl in combination with the experimental result of $D_A$ for
the same atom gives the most accurate limit for the S-PS
coupling constant to date.

We are grateful to Professor Eugene Commins for helpful
communications. We would like to thank Professors Daniel
Heinzen and David Weiss for information about the status
of their experiments. B. K. Sahoo is thankful to DAAD for
providing him with a scholarship. The calculations were
carried out on C-DAC's Param Padma supercomputer.

\end{document}